\documentclass[aps,prx,twocolumn,footinbib,floatfix]{revtex4-2}
\usepackage{graphicx}
\usepackage{indentfirst}
\usepackage{siunitx}
\usepackage{braket}
\usepackage{float}
\usepackage{amsmath}
\usepackage{physics}
\usepackage{amssymb}
\usepackage{CJK}
\usepackage{makecell}
\usepackage{esint}
\usepackage{color}
\usepackage[T1]{fontenc}
\usepackage{amsfonts}
\usepackage{footmisc}
\usepackage{scrextend}
\usepackage{multirow}
\usepackage[outdir=./]{epstopdf}

\usepackage[english]{babel}
\usepackage{url}
\usepackage{comment}
\usepackage{array}
\usepackage{subfiles}
\usepackage{tikz}
\usetikzlibrary{shapes,arrows}
\usepackage{mathrsfs}
\usepackage[mathscr]{eucal}
\usepackage{comment}
\usepackage{tabularx}
\usepackage{diagbox}

\usepackage[hyperfootnotes=false]{hyperref}
\usepackage{cleveref}
\definecolor{darkblue}{rgb}{0,0,0.5}
\hypersetup{
colorlinks=true,
linkcolor=black,
filecolor=blue,
citecolor=darkblue,  
urlcolor=black,
}

\def\be{\begin{equation}}
\def\ee{\end{equation}}
\def\ba{\begin{eqnarray}}
\def\ea{\end{eqnarray}}
\def\bal{\begin{equation}\begin{aligned}}
\def\eal{\end{aligned}\end{equation}}

\def\bp{\begin{pmatrix}}
\def\ep{\end{pmatrix}}

\usepackage{bm}

\newcommand{\calL}{{\cal L}}

\usepackage{pict2e,picture,graphicx}
\makeatletter
\DeclareRobustCommand{\Arrow}[1][]{%
\check@mathfonts
\if\relax\detokenize{#1}\relax
\settowidth{\dimen@}{$\m@th\rightarrow$}%
\else
\setlength{\dimen@}{#1}%
\fi
\sbox\z@{\usefont{U}{lasy}{m}{n}\symbol{41}}%
\begin{picture}(\dimen@,\ht\z@)
\roundcap
\put(\dimexpr\dimen@-.7\wd\z@,0){\usebox\z@}
\put(0,\fontdimen22\textfont2){\line(1,0){\dimen@}}
\end{picture}%
}
\makeatother

\usepackage[normalem]{ulem}
\usepackage{xr}
\begin{document}

\title{Variational Quantum Transduction}
\author{Pengcheng Liao${}^{1}$}

\author{Haowei Shi${}^{1}$}

\author{Quntao Zhuang${}^{1,2}$}
\email{qzhuang@usc.edu}

\affiliation{
${}^{1}$Ming Hsieh Department of Electrical and Computer Engineering, University of Southern California, Los
Angeles, California 90089, USA
\\
${}^{2}$Department of Physics and Astronomy, University of Southern California, Los
Angeles, California 90089, USA
}

\begin{abstract}
Quantum transducers are critical for quantum interconnect, enabling coherent signal transfer across disparate frequency domains. Beyond material and device advances, protocol design has become a powerful means to improve transduction. We introduce a variational quantum transduction (VQT) framework that employs variational tools from near-term quantum computing to systematically optimize protocol performance. 
As a variational quantum circuit framework, VQT is not plagued by known training issues such as barren plateau, because a small-scale problem is sufficient for substantial advantage and training only needs to be done once to configure a VQT system.
Maximizing the quantum information rate within this framework yields protocols that surpass all known schemes in their respective classes.
For non-adaptive protocols, VQT exceeds the performance envelopes of Gottesman-Kitaev-Preskill (GKP)-based and entanglement-assisted approaches. In the adaptive setting, VQT provides only a marginal improvement over Gaussian feedforward strategies, indicating that Gaussian adaptive transduction is already close to optimal.
%In the non-adaptive case, we show that the optimal protocol demonstrate GKP signatures in the low transmissivity region and transits towards Gaussian protocols towards the high transmissivity region. In the adaptive case, our optimization indicates that previously known adaptive quantum transduction scheme is close to optimal. 
With increasingly universal quantum control, VQT provides a systematic path toward optimal quantum transduction.

\end{abstract}
\maketitle

%\tableofcontents

\section{Introduction}
Quantum transduction (QT) aims to faithfully transfer quantum states between otherwise incompatible physical carriers, most notably between gigahertz superconducting circuits, which excel as quantum processors, and optical photons, which enable low-loss transmission over fiber~\cite{lauk2020perspectives,awschalom2021development,han2021microwave}. 
As such, transduction is a key primitive for quantum networks~\cite{acin2007entanglement,kimble2008quantum,wehner2018quantum}, distributed quantum sensing~\cite{zhang2021distributed,zhuang2018distributed,eldredge2018optimal,proctor2018multiparameter}, and distributed quantum computing~\cite{jiang2007distributed,monroe2014large}. 
Alongside rapid experimental progress in electro-optomechanical~\cite{bochmann2013nanomechanical,andrews2014bidirectional}, electro-optic~\cite{holzgrafe2020cavity}, and magneto-optic~\cite{hisatomi2016bidirectional} platforms, protocol design plays a central role in mitigating the performance bottlenecks of current devices, which often operate at sub-percent efficiencies and with added-noise occupations above a single photon~\cite{xu2024optomechanical}. 
%Throughout this work, we denote the transducer efficiency by~$\eta$.

Recent progress in protocol design can be broadly organized by the physical resources they exploit. 
First, \emph{feedforward} strategies, including adaptive Gaussian control~\cite{Zhang2018Quantum} and teleportation-based transduction~\cite{zhuang2025quantum}, can relax impedance-matching conditions and tolerate imperfect measurements. 
Second, \emph{engineered-environment and non-Gaussian} approaches, such as Gottesman–Kitaev–Preskill (GKP) environment-assisted schemes, can enable high-fidelity conversion well below the direct-conversion threshold~\cite{Wang2025Passive}. 
Third, \emph{entanglement assistance} within the microwave band can surpass efficiency--bandwidth limits via squeezer--coupler--antisqueezer architectures~\cite{shi2024overcoming}. 
Despite these advances, existing protocols are often analyzed with different performance metrics and resource assumptions, making it difficult to compare different schemes under a unified set of practical constraints.

In this work, we introduce variational quantum transduction (VQT), a framework that systematically searches for high-performance transduction protocols using variational quantum circuits (VQCs).
VQT optimizes the joint preparation of input states, ancillary resources, and decoding operations under explicit constraints, including bounds on energy per mode, the number of ancilla modes, and the availability of feedforward.
Although variational quantum circuit faces challenges such as barren plateau~\cite{mcclean2018barren,cerezo2021cost,zhang2024energy,larocca2025barren} in quantum computing applications, these issues do not arise here: transduction protocols involve only a small, finite number of modes, and the variational optimization needs be performed only once to configure the system, analogous to inference in classical machine learning.

To enable direct comparison across disparate protocol classes, we use the (single-letter) coherent information as the performance metric, which provides a computable lower bound on the quantum capacity. Within this framework, we uncover a clear operational distinction between non-adaptive and adaptive transduction. In the non-adaptive setting, VQT identifies protocols that substantially outperform both GKP-based environment-assisted schemes~\cite{Wang2025Passive} and intraband entanglement-assisted protocols~\cite{shi2024overcoming} under the same energy constraints. The optimized inputs moreover exhibit a clear transition with transmissivity: in the low-transmissivity regime ($\eta \lesssim 0.4$), the optimal signal and environment states are close to GKP states, indicating that non-Gaussian structure is the primary performance driver, whereas at larger~$\eta$ the optimal states become increasingly Gaussian and the remaining advantage is dominated by entanglement assistance.  In contrast, when adaptive feedforward is available, VQT yields only a modest improvement over Gaussian adaptive strategies~\cite{Zhang2018Quantum}, revealing that Gaussian adaptive transduction is already close to optimal.

\section{Variational Quantum Transduction Scheme}
\begin{figure}[t]
\centering
\includegraphics[width=0.95\linewidth]{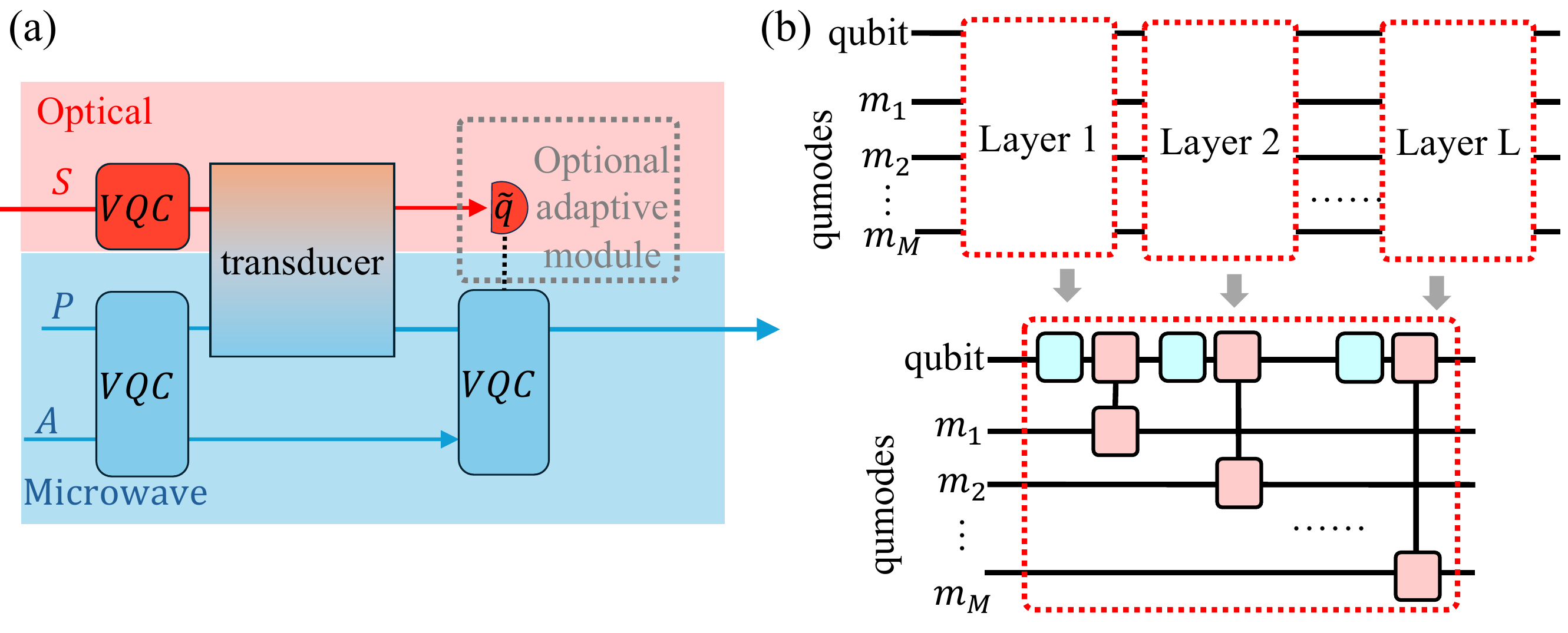}
\caption{(a) Variational transduction scheme. VQCs prepare the optical inputs $S$ and microwave inputs $(P,A)$ and apply a final decoding VQC after the transducer. 
(b) Schematic of a multi-layer ECD variational circuit. Each layer consists of qubit–qumode conditional-displacement operations (pink) and single-qubit rotations (blue), providing universal hybrid control.
}
\label{fig:variational_NonAdaptive}
\label{fig:ECD-circuit}
\label{fig:variational_Adaptive}
\end{figure}

Quantum transduction converts quantum information between bosonic modes $\hat{a}$ and $\hat{b}$ at different frequencies. We can generally model such a transduction process as a beamsplitter interaction $\hat{a}\hat{b}^\dagger$ with linear input output relation
\be 
\hat{a}^\prime=\sqrt{\eta}\hat{a}+\sqrt{1-\eta}\hat{b},
\hat{b}^\prime=\sqrt{\eta}\hat{b}-\sqrt{1-\eta}\hat{a},
\ee 
between the two input modes and two output modes~\cite{shi2024overcoming,Wang2025Passive}, where $\eta$ is the conversion efficiency. While the enhancement of $\eta$ towards unity relies on material engineering and device fabrication, quantum protocols can engineer the input quantum states of $\hat{a}$ and $\hat{b}$, such as the GKP-based approach~\cite{Wang2025Passive}, or introduce additional ancilla modes with pre- and post-processing, such as the entanglement-assisted scheme~\cite{shi2024overcoming}.
As shown in Fig.~\ref{fig:variational_NonAdaptive}(a), the proposed VQT scheme adopts VQCs to systematically optimize the quantum transduction protocol. While conventionally adopted for quantum computing, VQCs provide a flexible and hardware-compatible framework for designing transduction protocols under realistic constraints. We will take the example of optical to microwave transduction to further illustrate the scheme.

The VQT framework employs three basic variational components: (i) a circuit that prepares the optical input $S$, (ii) a circuit preparing the microwave input $(P,A)$, and (iii) a joint decoder acting on $(P,A)$ after the transducer. 
%In this work, VQCs serve as the parameterized family over which we optimize the coherent information, enabling the discovery of nontrivial encoding and decoding transformations that incorporate non-Gaussian resources, feedforward, and entanglement assistance. %This section introduces the control primitives, circuit architecture, and the distinction between non-adaptive and adaptive VQT schemes.
In addition, an optional adaptive module can be adopted, which performs measurement on the output of $S$ to feedforward to the VQC to produce the final output. For simplicity, we consider a displacement of $D(\tilde{q})$ conditional on the homodyne measurement result $\tilde{q}$. Due to additional VQC components before and after the displacement, such a construction allows general unitary in the feedforward operation.

The proposed VQC therefore has two modes of operation.
In the non-adaptive setting, we delete the optional adaptive module, thus no mid-circuit measurement outcomes influence subsequent operations. The ansatz is expressive enough to represent arbitrary input states and decoding transformations allowed by energy constraints, thereby fully exploiting non-Gaussianity and entanglement assistance where beneficial.
In contrast, the adaptive protocol can approximate an arbitrary feedforward-controlled unitary. The resulting architecture fully leverages entanglement assistance, non-Gaussian states, and measurement-based adaptivity.

% \begin{figure}[htbp]
% \centering
% \includegraphics[width=0.9\linewidth]{variational_NonAdaptive.pdf}
% \caption{Variational non-adaptive transduction scheme. VQCs prepare the joint optical inputs $(R,S)$ and microwave inputs $(P,A)$ and apply a final decoding VQC after the transducer. No feedforward is used.}
% \label{fig:variational_NonAdaptive}
% \end{figure}
% \begin{figure}[htbp]
% \centering
% \includegraphics[width=0.98\linewidth]{Variational_Adaptive.pdf}
% \caption{Variational adaptive transduction scheme. A homodyne measurement on the optical output conditions a displacement $D(\tilde{q})$ on the microwave mode. VQCs before and after the feedforward step provide a universal adaptive variational decoder.}
% \label{fig:variational_Adaptive}
% \end{figure}

%Specifically, we employ echoed conditional-displacement (ECD) gates~\cite{Eickbusch2022Fast} and conditional-NOT displacement gates~\cite{Diringer2023Conditional}, which are available across multiple platforms, to realize universal hybrid control and train variational encoders and decoders. 

Among several universal hybrid-control architectures, we adopt the echoed conditional-displacement (ECD) gate~\cite{Eickbusch2022Fast} as a representative hardware-native primitive for the VQCs. The ECD gate applies opposite phase-space displacements conditioned on a qubit state,
\begin{equation}
\hat{U}_{\rm ECD}(\beta)=\hat{D}(\beta)\otimes\ket{1}\!\bra{0}
+\hat{D}(-\beta)\otimes\ket{0}\!\bra{1},
\end{equation}
and, together with single-qubit rotations, enables expressive control over hybrid qubit-bosonic systems. In our implementation, each variational circuit consists of $L$ repeated layers of ECD interactions and single-qubit rotations, as illustrated in Fig.~\ref{fig:ECD-circuit}(b).

% Among several equivalent universal hybrid-control frameworks, we adopt the echoed conditional-displacement (ECD)~\cite{Eickbusch2022Fast} gate 
% \begin{equation}
% \hat{U}_{\rm ECD}(\beta)=\hat{D}(\beta)\otimes\ket{1}\!\bra{0}
% +\hat{D}(-\beta)\otimes\ket{0}\!\bra{1},
% \end{equation}
% where the single-mode displacement is
% $
% \hat{D}(\beta)=\exp(\beta \hat{a}^\dagger-\beta^* \hat{a}).
% $
% %with the canonical quadratures defined as$\hat{q}=(\hat{a}+\hat{a}^\dagger)/\sqrt{2},\ \hat{p}=(\hat{a}-\hat{a}^\dagger)/(\sqrt{2}i).$
% Together with single-qubit rotations,
% % \begin{equation}
% % \hat{U}_{\rm R}(\xi,\phi)=\exp[-i\xi/2(\cos\phi\,\hat{\sigma}^x+\sin\phi\,\hat{\sigma}^y)],
% % \end{equation}
% the ECD gate forms a universal gate set for hybrid registers consisting of qubits and bosonic modes~\cite{Eickbusch2022Fast,zhang2024energy}. 

% We denote the multimode displacement operator as
% \begin{equation}
% \hat{D}(\bm{x}) \equiv \bigotimes_{m=1}^{M} \hat{D}(q_m+i p_m),
% \end{equation}
% where $\bm x=(q_1,p_1,\dots,q_M,p_M)$ specifies the phase-space displacements applied to each mode. 

% For an $M$-mode system, each variational circuit $\hat{U}(\boldsymbol{\theta})$ is composed of $L$ repeated layers of ECD interactions and single-qubit rotations, as illustrated in Fig.~\ref{fig:ECD-circuit}(b). These layers constitute the VQT building block from which encoders and decoders are constructed.

\section{Performance comparison}
To benchmark the performance of variational quantum transduction, besides direct transduction with the bare transducer, we compare against three representative baseline strategies that leverage distinct physical resources: intraband entanglement assistance (EA) via two-mode squeezing (TMS)~\cite{shi2024overcoming}, GKP-based environment engineering~\cite{Wang2025Passive}, and adaptive quantum transduction protocol~\cite{Zhang2018Quantum}. Details are provided in Appendix.
These protocols capture the best known approaches in their respective resource regimes and serve as meaningful reference points for VQT. 

Different performance metrics have been used in prior work, including entanglement fidelity~\cite{Wang2025Passive} and effective transmissivity enhancement~\cite{shi2024overcoming}. 
To enable a unified comparison across various protocols, we adopt the \emph{coherent information} as the performance measure throughout this work.

More precisely, let $\mathcal{N}:{\cal B}(\mathcal{H}_S)\!\to\!{\cal B}(\mathcal{H}_P)$ denote a quantum channel mapping the input mode $S$ to the output mode~$P$.  
For an input state $\rho_S\in{\cal D}(\mathcal{H}_S)$, then the \emph{coherent information} of $\rho_S$ through $\mathcal{N}$ is defined as
\be 
  I_c(\rho_S,\mathcal{N})
  := S(\rho_P)-S(\rho_{RP}),
\ee 
where $S(\cdot)$ is the von Neumann entropy. Here, we introduce reference $R$ to purify the input for evaluation purpose. $RS$ is in a pure state $\ket{\psi}$ such that 
$\Tr_R\!\left[\ket{\psi}\!\bra{\psi}_{RS}\right]=\rho_S$. And the joint output state
$\rho_{RP}=(\mathcal{I}_R\!\otimes\!\mathcal{N})\!\left[\ket{\psi}\!\bra{\psi}_{RS}\right]$ and the reduced state $\rho_P=\Tr_R(\rho_{RP})$.

\begin{figure}[t]
    \centering
    \includegraphics[width=0.95\linewidth]{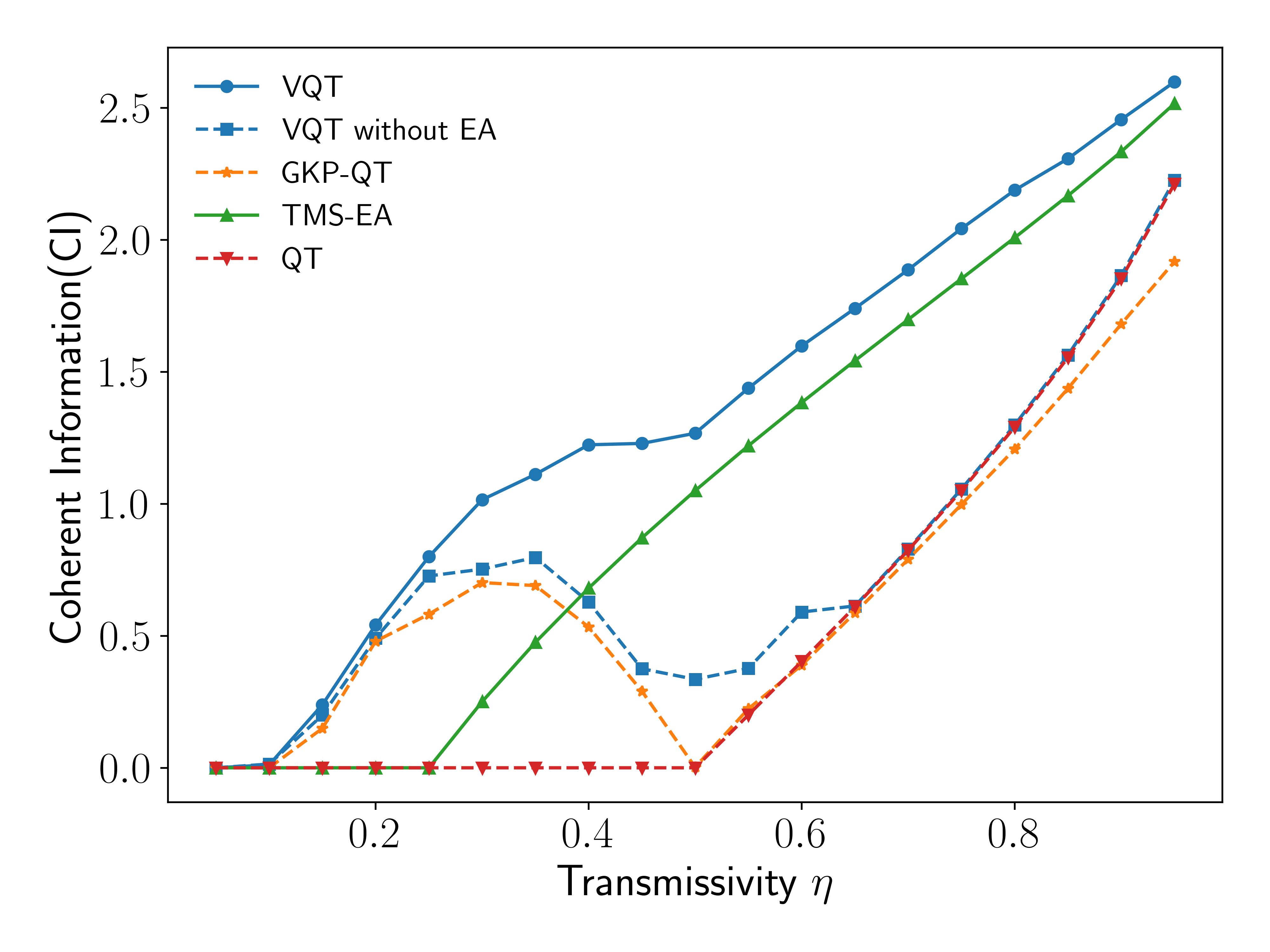}
    \caption{Coherent information achieved by various non-adaptive transduction protocols with energy constraints $n_S=n_P=2$. The fully variational entanglement-assisted scheme (VQT) yields the highest performance across all $\eta$. The single-mode variant (VQT without EA) follows a similar trend but with reduced values, highlighting the benefit of entanglement assistance. The GKP-QT model, using finite-energy GKP state for both for $S$ and  $P$, matches VQT (with and without EA) at low $\eta$ and deviates at higher transmissivity. The TMS-EA protocol shows a threshold-like onset and approaches VQT at high $\eta$, but remains suboptimal overall.
}
    \label{fig:variational_NonAdaptive_CI}
\end{figure}

The \emph{single-letter coherent information} of the channel is obtained by maximizing over all input states,
\begin{equation}
  I_c(\mathcal{N}) := \max_{\rho_S} I_c(\rho_S,\mathcal{N}).
\end{equation}
The \emph{quantum capacity} of the channel is the maximum asymptotic rate of reliable quantum communication, given by the regularized coherent information~\cite{lloyd1997capacity,shor2002capacity,devetak2005private}:
$
  Q(\mathcal{N})
  := \lim_{n\to\infty}
  \frac{1}{n}
  \max_{\rho_{S^n}}
  I_c\!\left(\rho_{S^n},\mathcal{N}^{\otimes n}\right).
$
Because $I_c(\mathcal{N})$ provides a computable lower bound on $Q(\mathcal{N})$, it serves as a practical and meaningful performance metric for transduction.

In VQT, both the channel~$\mathcal{N}$ and the input state~$\rho_S$ depend on the choice of variational quantum circuits.  
We therefore define the optimization target as
$
    I_c := \max_{\rm VQC}\, I_c(\rho_S,\mathcal{N})
$
and use this quantity as the objective function for VQC training, under the energy constraints on the input modes to the transducer. More details of the evaluation are in Appendix.
The optimized coherent information achieved by VQT, along with comparisons to the baseline protocols, is presented and analyzed in the following subsections.
To understand the resource needed, we also consider a VQT without the microwave ancilla $A$, which we refer to as `VQT-without-EA'.

\section{Non-Adaptive Protocols}
\begin{figure}[t]
    \centering
    \includegraphics[width=0.99\linewidth]{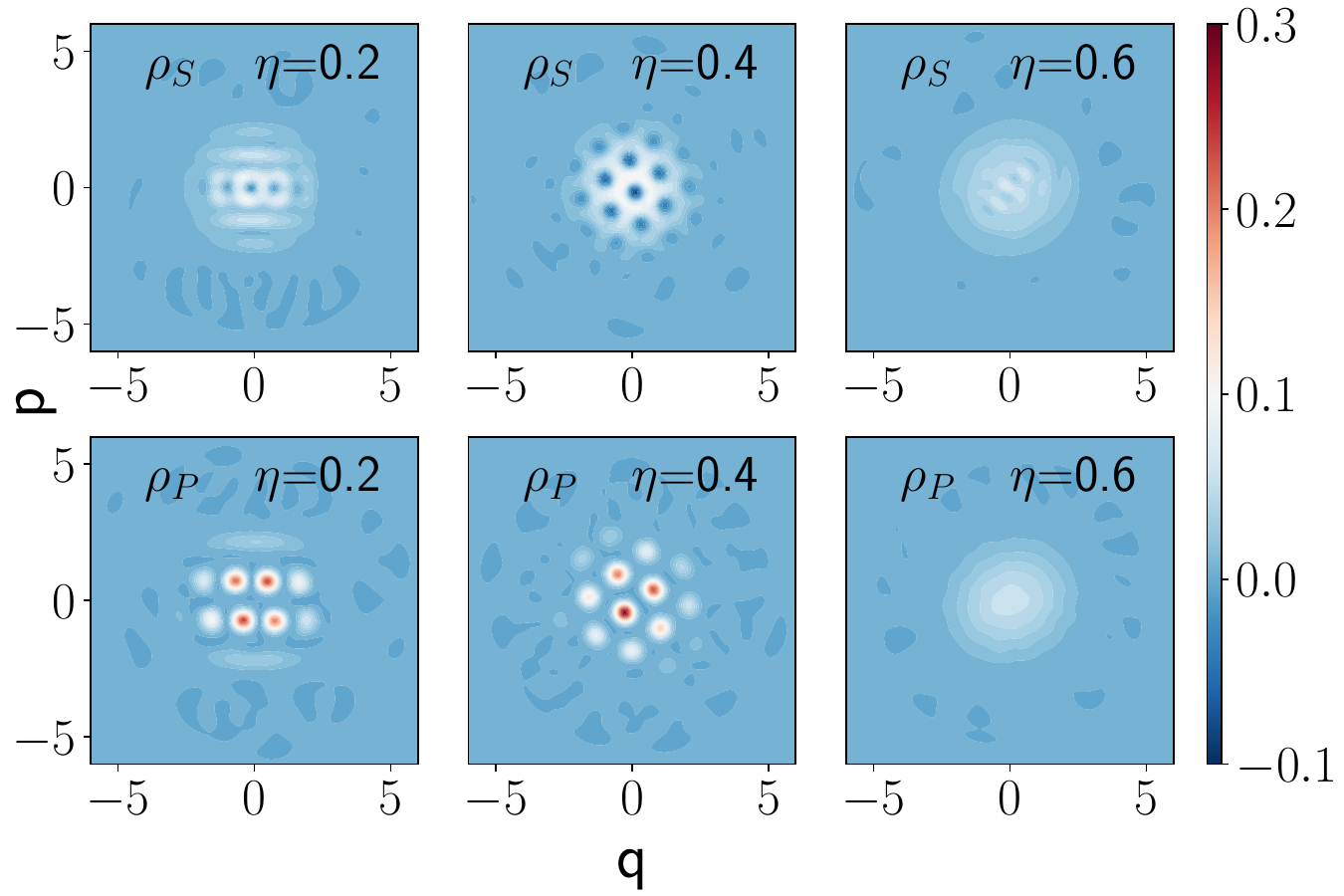}
    \caption{Wigner functions of the optimized input states for the VQT protocol at different transmissivities~$\eta$.  
For small $\eta (\lesssim0.4)$ , the optimal signal state of $\rho_S$ and $\rho_P$  exhibit the characteristic lattice structure of a GKP-like state. 
As $\eta$ increases, both $\rho_S$ and $\rho_P$ gradually lose their non-Gaussian structure and approach Gaussian states.}
    \label{fig:wigner_function_nonadaptive_ECD_MM}
\end{figure}

The coherent information achieved by different non-adaptive protocols is shown in Fig.~\ref{fig:variational_NonAdaptive_CI}, where the average photon numbers in modes $S$ and $P$ are restricted to $n_S=n_P=2$. Across the entire range of $\eta$, the fully variational entanglement-assisted protocol (VQT) yields the highest coherent information, increasing smoothly and monotonically. 
To gain insight into the behavior of the VQT protocol, we analyze the optimized variational circuits and examine the corresponding input states. As shown in Fig.~\ref{fig:wigner_function_nonadaptive_ECD_MM}, for small $\eta$, the optimal states of $S$ and $P$ are close to GKP states. See the Appendix for more details. 
As $\eta$ increases, both $S$ and $P$ gradually approach Gaussian states. 

%Motivated by these observations, we construct a reduced variational model in which the input states are restricted to structured entangled families. For the optical registers $(R,S)$, we parameterize the input as an entangled approximate GKP state, \begin{equation}    \ket{\psi}_{RS} \sim \sum_{i=1}^{d} a_i\, \ket{i}_R \ket{{\rm GKP}_{\rm approx}(d,i)}_{S},\end{equation} where $\ket{i}$ denotes the $i$th Fock basis state and $\ket{{\rm GKP}_{\rm approx}(d,i)}$ is a finite-energy approximation of a GKP qudit at logical index~$i$. Similarly, for the microwave registers $(P,A)$, we parameterize the joint state as an entangled squeezed-coherent superposition, \begin{equation} \ket{\psi}_{PA} \sim \sum_{i=1}^{d} b_i\, \ket{\alpha_i,\zeta_i}_{P}\ket{i}_A,\end{equation} where $\ket{\alpha_i,\zeta_i}$ is a squeezed coherent state and $\ket{i}_A$ is the Fock basis of the ancilla.  Optimizing over the parameters $\{a_i,b_i,\alpha_i,\zeta_i\}$ and the decoding variational circuit yields the curve labeled GKP-EA in Fig.~\ref{fig:variational_NonAdaptive_CI}. 

Comparing VQT with VQT-without-EA,  we observe that entanglement assistance offers little improvement when $\eta$ is small, but becomes increasingly beneficial as $\eta$ grows. This behavior highlights that entanglement is most effective in the high-transmissivity regime, while its impact remains modest at low $\eta$.

Conversely, the advantage of VQT over TMS-EA, or VQT-without-EA over the QT capacity, is most pronounced when $\eta$ is small. As $\eta$ increases, this gap steadily narrows and in some cases vanishes entirely—for instance, VQT-without-EA approaches the QT channel capacity once $\eta > 0.6$. This trend indicates that non-Gaussian resources provide the greatest benefit in the low-transmissivity regime, while offering limited improvement when the channel becomes sufficiently transparent.

%%this repeats the conclusion
%In summary, the numerical results show that the VQT scheme consistently achieves the highest coherent information across the full range of transmissivities by combining entanglement assistance and non-Gaussian structure. Non-Gaussianity is the key performance driver at low $\eta$, whereas entanglement becomes the dominant resource as transmissivity increases.

\begin{figure}[t]
    \centering
    \includegraphics[width=0.95\linewidth]{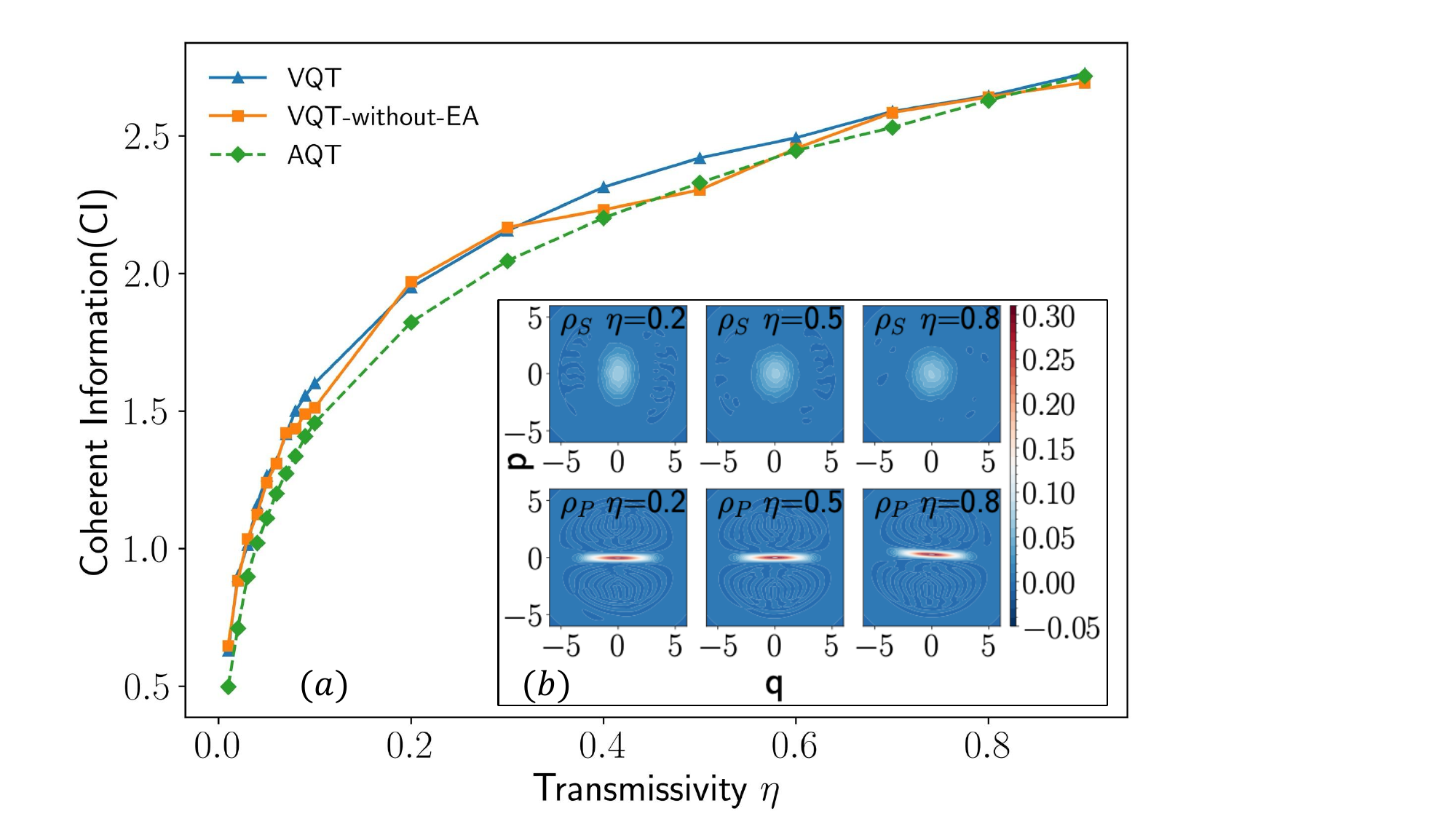}
    \caption{(a)~Coherent information of adaptive transduction protocols as a function of transmissivity~$\eta$. The variational schemes (VQT and VQT-without-EA) exhibit nearly identical performance, indicating that feedforward effectively removes the benefits of additional ancilla modes. The adaptive Gaussian protocol (AQT) closely tracks both variational results.
    (b) Wigner functions of the optimized input states for the VQT protocol. Across all transmissivities~$\eta$, the optimal input to $S$ is a squeezed thermal state, while the optimal input to $P$ is a squeezed vacuum. Neither entanglement nor non-Gaussian features appear in the optimized inputs.
}
    \label{fig:variational_Adaptive_CI_Wigner}
\end{figure}

\section{Adaptive Protocols}
For numerical convenience in evaluating coherent information in the adaptive setting, we replace homodyne measurement followed by feed-forward with a SUM gate, since both procedures induce the same completely positive trace-preserving (CPTP) map on subsystem $P$. More details and proof are given in Appendix Sec.II.

Figure~\ref{fig:variational_Adaptive_CI_Wigner} shows the coherent information achieved by several adaptive transduction protocols. The fully variational multimode strategy (VQT) yields the highest values, but only by a small margin. The single-mode variational scheme (VQT-without-EA) nearly overlaps with VQT, demonstrating that adaptive feedforward largely eliminates the benefit of ancillary modes. The previously proposed adaptive Gaussian protocol~(AQT) performs comparably, tracking both variational strategies at moderate and large~$\eta$. This behavior suggests that, once feedforward is available, Gaussian processing is already close to optimal and only minor improvements can be gained through additional variational freedom.

Inspection of the optimized VQT inputs (Fig.~\ref{fig:variational_Adaptive_CI_Wigner}) further supports this conclusion. Across the full range of~$\eta$, the optimal state in mode~$S$ is a squeezed thermal state and the optimal state in~$P$ is a squeezed vacuum. The optimized $P\!A$ register contains no entanglement, nor does either subsystem exhibit non-Gaussian structure, indicating that neither resource meaningfully improves performance in the adaptive setting.

We offer an intuitive explanation for why entanglement provides limited benefit in the variational adaptive setting. Since the transducer interaction is modeled as a beam splitter, the output modes $S$ and $P$ both retain partial information about the input state. In the moderate transmissivity regime, this shared information creates a fundamental limitation: if both modes contained enough information to perfectly reconstruct the original state in $S$, the process would effectively generate two copies, violating the no-cloning theorem~\cite{wootters1982single}. 

In the non-adaptive setting, this limitation can be mitigated because the beam splitter not only distributes signal information, but also transfers appropriately engineered noise, introduced via non-Gaussianity or entanglement assistance—into the $S$ mode. This contamination suppresses the residual information in $S$, allowing the $P$ mode to better approximate a faithful reconstruction without contradicting no-cloning constraints.

In contrast, in the adaptive setting the signal mode is measured, eliminating its remaining quantum information. As a result, there is no need to suppress information in $S$ via added non-Gaussianity or entanglement: measurement already removes the potential cloning conflict. Consequently, a Gaussian feedforward strategy becomes nearly optimal, explaining the diminishing advantage of non-Gaussian or entanglement-assisted resources in the adaptive regime.

\section{Discussions}
In this work, we introduced the VQT framework, in which variational quantum circuits are used to optimize transduction performance as quantified by coherent information. In the non-adaptive setting, we observe that non-Gaussianity and entanglement assistance enhance performance in complementary transmissivity regimes. The combined VQT scheme, which incorporates both resources, achieves the highest overall performance and outperforms all baseline protocols by a significant margin. 
%In contrast, in the adaptive setting the advantage of non-Gaussianity and entanglement largely disappears: VQT offers only minor improvements beyond existing protocols.
Our protocol can be interpreted as a quantum communication protocol for two-way quantum channels or quantum interactions~\cite{bennett2003capacities,zanardi2000entangling,childs2006two}, with single-sided entanglement assistance. Refs.~\cite{Wang2025Passive,shi2024overcoming} indicate that in the energy unconstrained case, the capacity for the beamsplitter channel relevant in transduction can be unbounded; the energy-constrained case is largely unstudied, and our work reveals an achievable lower bound interpolating between the two approaches.

%, and the Gaussian, entanglement-free AQT scheme is already nearly optimal. %This behavior reflects the fact that measurement and feedforward effectively eliminate the benefits that non-Gaussian structure and entanglement provide in the non-adaptive regime.

%Our results also raise several open questions and point toward promising research directions. A key unresolved issue is the role of energy constraints. In the infinite-energy limit, either entanglement assistance or non-Gaussian encoding can, in principle, boost effective transduction efficiency to unity; however, it remains unclear whether the optimality transitions observed here persist, vanish, or shift under asymptotically large squeezing. Understanding how these resource advantages scale with energy remains an important avenue for future investigation.

\begin{acknowledgements}
The project is supported by DARPA (HR0011-24-9-0362) and NSF (CCF-2240641, 2350153) and ONR (N00014-23-1-2296). Q.Z. also acknowledges support from Google.
\end{acknowledgements}

\appendix
\section{Preliminaries}

\subsection{Basics}

Let $\hat a$ and $\hat a^\dagger$ denote the annihilation and creation operators of a
single bosonic mode. The dimensionless canonical quadratures are
\begin{equation}
\hat q=\frac{\hat a+\hat a^\dagger}{\sqrt2},
\qquad
\hat p=\frac{\hat a-\hat a^\dagger}{i\sqrt2},
\end{equation}
which satisfy $[\hat q,\hat p]=i$ (we set $\hbar=1$). The single-mode phase rotation
operator is
\begin{equation}
\hat R(\theta)=\exp\!\left(-i\theta\,\hat a^\dagger\hat a\right),
\end{equation}
corresponding to a rotation by angle $\theta\in\mathbb R$ in phase space. The
single-mode squeezing operator is
\begin{equation}
\hat S(\zeta)
=\exp\!\left[
\frac{1}{2}\left(\zeta^* \hat a^2-\zeta\,\hat a^{\dagger 2}\right)
\right],
\end{equation}
where the complex squeezing parameter $\zeta=re^{i\phi}$ has magnitude $r\ge 0$ and
phase $\phi\in\mathbb R$.

\subsection{Gottesman--Kitaev--Preskill (GKP) states}

Gottesman--Kitaev--Preskill (GKP) codes encode a finite-dimensional qudit (a
$d$-level system) into a single bosonic mode by specifying a lattice in phase
space~\cite{GKP2001}. Define the displacement operator
\begin{equation}
\hat T(\xi)
=\exp\!\left(i\,\xi_p \hat q - i\,\xi_q \hat p\right),
\qquad
\xi=(\xi_q,\xi_p)\in\mathbb R^2 .
\end{equation}
A (single-mode) GKP code is associated with a rank-two lattice
$\Lambda\subset\mathbb R^2$ generated by primitive vectors
$\bm v=(v_q,v_p)$ and $\bm u=(u_q,u_p)$. The corresponding displacement operators
commute 
\begin{equation}
\hat T(\bm v)\hat T(\bm u)=\hat T(\bm u)\hat T(\bm v),
\end{equation}
when
\begin{equation}
\bm v^{\sf T}\Omega\,\bm u
= v_q u_p - v_p u_q \in 2\pi\mathbb Z,
\qquad
\Omega=
\begin{pmatrix}
0 & 1\\
-1& 0
\end{pmatrix}.
\end{equation}
Choosing the primitive unit cell to have symplectic area $2\pi d$ yields a
$d$-dimensional logical subspace.

Equivalently, one may specify the code as $\mathcal C_{d,S}$, where $S$ is the
normalized orientation matrix
\begin{equation}
S=\frac{1}{\sqrt{2\pi d}}
\begin{pmatrix}
\bm v\\[2pt]
\bm u
\end{pmatrix}
=
\frac{1}{\sqrt{2\pi d}}
\begin{pmatrix}
v_q & v_p\\
u_q & u_p
\end{pmatrix},
\label{eq:orientation_matrix}
\end{equation}
so that $\det(S)=1$.

The stabilizer group is generated by the commuting displacement operators
\begin{equation}
\hat S_X=\hat T(\bm v),
\qquad
\hat S_Z=\hat T(\bm u),
\end{equation}
and the codespace is the joint $+1$ eigenspace of $\hat S_X$ and $\hat S_Z$:
\begin{equation}
\mathcal C_{d,S}
=
\operatorname{span}\Bigl\{\ket{\psi}\,:\,
\hat S_X\ket{\psi}=\ket{\psi},\ \hat S_Z\ket{\psi}=\ket{\psi}
\Bigr\}.
\end{equation}

For the square lattice (normalized orientation $S=I_2$), one can choose
$\bm v=(\sqrt{2\pi d},0)$ and $\bm u=(0,\sqrt{2\pi d})$. The ideal logical basis
states may be written in the position basis as Dirac combs,
\begin{equation}
\ket{i}_L \propto \sum_{s\in\mathbb Z}
\ket{q=\sqrt{\frac{2\pi}{d}}\,(ds+i)},
\qquad
i\in\{0,1,\ldots,d-1\}.
\end{equation}

Ideal GKP codewords have infinite mean photon number. A commonly used finite-energy
approximation is obtained by applying a Gaussian envelope in Fock space~\cite{GKP2001,noh2018quantum},
\begin{equation}
\ket{i_\Delta}_L
:= \mathcal N_{\Delta,i}\,e^{-\Delta^2 \hat n}\ket{i}_L,
\end{equation}
where $\mathcal N_{\Delta,i}$ is a normalization constant and
$\hat n=\hat a^\dagger\hat a$ is the photon number operator.
More general lattice geometries can be obtained by applying a single-mode symplectic
transformation (rotation--squeezing--rotation) to the square-lattice state,
\begin{equation}
\ket{(d,i)}_{\mathrm{GKP}}
=
\hat R(\phi_2)\,\hat S(r)\,\hat R(\phi_1)\,\ket{i_\Delta}_L.
\label{eq:GKP_finite_n_general}
\end{equation}

\subsection{Pure loss channel}

The pure-loss channel $\mathcal{L}_\eta$ can be modeled as a beamsplitter of transmissivity $\eta\in[0,1]$ that couples the input signal mode $\hat a_{\rm in}$ to an environmental mode $\hat e$ prepared in the vacuum state $\ket{0}$, followed by tracing out the environment:
\be
\hat{a}_{\rm out} = \sqrt{\eta}\,\hat{a}_{\rm in} + \sqrt{1-\eta}\,\hat{e}.
\ee
Its unconstrained quantum capacity is
\be
Q(\mathcal{L}_\eta) = \max\left\{0,\log_2\frac{\eta}{1-\eta}\right\}.
\label{eq:Q_pure_loss}
\ee
With an energy constraint $\langle \hat n\rangle \le n$ on the input state, the quantum capacity is~\cite{noh2018quantum}
\be
Q(\mathcal{L}_\eta,n) = \max\left\{0,\, g(\eta n)- g\bigl((1-\eta)n\bigr)\right\},
\label{eq:Q_pure_loss_energy_constraint}
\ee
where
\be
g(x) = (x+1)\log_2(x+1) - x\log_2(x).
\ee

\subsection{Additive noise channel}

The single-mode additive noise (random-displacement) channel $\mathcal{A}_{\sigma^2}$ is defined by
\be
\mathcal{A}_{\sigma^2}(\rho)
=
\iint \frac{d^{2}\alpha}{\pi\,\sigma^2}\,
e^{-|\alpha|^{2}/\sigma^2}\,
D(\alpha)\,\rho\,D^{\dagger}(\alpha),
\ee
where $D(\alpha)=\exp\!\left(\alpha \hat a^\dagger-\alpha^* \hat a\right)$ is the displacement operator. With this parametrization, the added noise has equal variance in the two quadratures,
\be
\sigma_q^2=\sigma_p^2=\sigma^2.
\ee

Under an energy constraint $\langle \hat n\rangle\le n$, a lower bound on the quantum capacity is~\cite{holevo2001evaluating, noh2020enhanced}
\be
Q(\mathcal{A}_{\sigma^2,n})
\ge
g\!\bigl(n+\sigma^{2}\bigr)
- g\!\left(\frac{D' + \sigma^{2} - 1}{2}\right)
- g\!\left(\frac{D' - \sigma^{2} - 1}{2}\right),
\ee
where
\be
D' \equiv \sqrt{\,\bigl(2n + \sigma^{2} + 1\bigr)^{2} - 4n\,(n+1)},
\ee
and this lower bound is achieved by thermal states.

\section{Equivalence of Homodyne\,$+$\,Feed-Forward with a SUM Gate}
In this section, we prove homodyne measurement followed by feed-forward  induce the same completely positive trace-preserving (CPTP) map on subsystem $P$ as with SUM gate.  A detailed derivation is given below.

\subsection{Two-mode preliminaries.}
Consider two bosonic modes $A$ and $B$ with annihilation (creation) operators $\hat a$ ($\hat a^\dagger$) and $\hat b$ ($\hat b^\dagger$), satisfying
\begin{equation}
[\hat a,\hat a^\dagger]=[\hat b,\hat b^\dagger]=1.
\end{equation}
With $\hbar=1$, define the canonical quadratures
\begin{align}
  \hat q_A=\frac{\hat a+\hat a^\dagger}{\sqrt2}, \qquad
  \hat p_A=\frac{\hat a-\hat a^\dagger}{i\sqrt2},\\
  \hat q_B=\frac{\hat b+\hat b^\dagger}{\sqrt2}, \qquad
  \hat p_B=\frac{\hat b-\hat b^\dagger}{i\sqrt2},
\end{align}
so that
\begin{equation}
[\hat q_A,\hat p_A]=i,\qquad [\hat q_B,\hat p_B]=i,\qquad [\hat q_A,\hat p_B]=0.
\end{equation}
A $q$-quadrature displacement on mode $\alpha\in\{A,B\}$ is
\begin{equation}\label{eq:DQ}
  \hat D_Q^{\alpha}(q)\;=\;\exp\!\bigl(-i\,q\,\hat p_{\alpha}\bigr).
\end{equation}
Acting on position eigenstates $\hat q_\alpha\ket{q_1}_\alpha=q_1\ket{q_1}_\alpha$, it satisfies
\begin{equation}
  \hat D_Q^{\alpha}(q)\ket{q_1}_\alpha=\ket{q_1+q}_\alpha .
\end{equation}

\subsection{SUM Gate}
For coupling strength $\lambda=1$, define the SUM gate
\begin{equation}\label{eq:SUM}
  \hat U_{\mathrm{SUM}} \;=\; \exp\!\bigl(-i\,\hat q_A\,\hat p_B\bigr).
\end{equation}
In the Heisenberg picture, it acts as
\begin{equation}
  \hat q_A\mapsto \hat q_A,\qquad
  \hat p_A\mapsto \hat p_A-\hat p_B,\qquad
  \hat q_B\mapsto \hat q_B+\hat q_A,\qquad
  \hat p_B\mapsto \hat p_B.
\end{equation}

Since $\hat q_A$ is self-adjoint, the spectral theorem gives
\begin{equation}
  \hat q_A=\int_{-\infty}^{\infty} dq\; q\,\ket{q}_A\,{}_A\!\bra{q}.
\end{equation}
Therefore, $\hat U_{\mathrm{SUM}}$ admits the resolution
\begin{align}
  \hat U_{\mathrm{SUM}}
  &= e^{-i\,\hat q_A\hat p_B}\notag\\
  &= \sum_{n=0}^{\infty}\frac{(-i\hat p_B)^n}{n!}\,\hat q_A^{\,n}\notag\\
  &= \sum_{n=0}^{\infty}\frac{(-i\hat p_B)^n}{n!}\,
     \int_{-\infty}^{\infty} dq\, q^n\,\ket{q}_A\,\!\bra{q}\notag\\
  &= \int_{-\infty}^{\infty} dq\; e^{-i\,q\,\hat p_B}\;\ket{q}_A\bra{q}\notag\\
  &= \int_{-\infty}^{\infty} dq\; \ket{q}_A \bra{q} \;\otimes\;\hat D_Q^{B}(q).
  \label{eq:SUM-resolved}
\end{align}

\subsection{Scenario A: Homodyne\,$+$\,Feed-Forward}
Let $\rho_{AB}$ be an arbitrary two-mode state. The (unnormalized) post-measurement state conditioned on outcome $q$ from an ideal homodyne measurement of $\hat q_A$ is
\begin{equation}
  \tilde\rho_{AB|q}
  \;=\;
  \bigl(\ket{q}_A\,{}_A\!\bra{q}\otimes\hat{\openone}_B\bigr)\,
  \rho_{AB}\,
  \bigl(\ket{q}_A\,{}_A\!\bra{q}\otimes\hat{\openone}_B\bigr).
\end{equation}
Apply feed-forward on mode $B$ using \eqref{eq:DQ}:
\begin{equation}
  \tilde\rho_{AB|q}'
  \;=\;
  (\hat{\openone}_A\otimes \hat D_Q^{B}(q))\,
  \tilde\rho_{AB|q}\,
  (\hat{\openone}_A\otimes \hat D_Q^{B\dagger}(q)).
\end{equation}
Discard the classical record (average over $q$) and trace out subsystem $A$:
\begin{align}
  \mathcal E_{\mathrm{hom+ff}}[\rho_{AB}]
  &:=\int_{-\infty}^{\infty} dq\;
     \operatorname{Tr}_A\!\bigl[\tilde\rho_{AB|q}'\bigr]\notag\\
  &=\int_{-\infty}^{\infty} dq\;
     \hat D_Q^{B}(q)\,
     {}_{A}\!\bra{q}\,\rho_{AB}\,\ket{q}_{A}\,
     \hat D_Q^{B\dagger}(q).
  \label{eq:map-hom}
\end{align}
Equation \eqref{eq:map-hom} is a Kraus decomposition of the induced CPTP map on subsystem $B$.

\subsection{Scenario B: SUM Gate Then Trace}
Alternatively, apply $\hat U_{\mathrm{SUM}}$ and then trace out $A$:
\begin{align}
  \mathcal E_{\mathrm{SUM+tr}}[\rho_{AB}]
  &:=\operatorname{Tr}_A\!\Bigl[
       \hat U_{\mathrm{SUM}}\,
       \rho_{AB}\,
       \hat U_{\mathrm{SUM}}^\dagger
     \Bigr]\notag\\[2pt]
  &=\int_{-\infty}^{\infty} dq\;
      {}_{A}\!\bra{q}\,
      \hat U_{\mathrm{SUM}}\,
      \rho_{AB}\,
      \hat U_{\mathrm{SUM}}^\dagger
      \ket{q}_{A},
  \label{eq:traceA}
\end{align}
where we inserted $\hat{\openone}_A=\int dq\,\ket{q}_A\!\bra{q}$ in the partial trace.
Using \eqref{eq:SUM-resolved},
\begin{align}
  {}_{A}\!\bra{q}\,
  \hat U_{\mathrm{SUM}}\,
  \rho_{AB}\,
  \hat U_{\mathrm{SUM}}^\dagger
  \ket{q}_{A}
  &=
  \hat D_Q^{B}(q)\,
  {}_{A}\!\bra{q}\,\rho_{AB}\,\ket{q}_{A}\,
  \hat D_Q^{B\dagger}(q).
\end{align}
Substituting into \eqref{eq:traceA} yields
\begin{equation}
  \mathcal E_{\mathrm{SUM+tr}}[\rho_{AB}]
  \;=\;
  \int_{-\infty}^{\infty} dq\;
  \hat D_Q^{B}(q)\,
  {}_{A}\!\bra{q}\,\rho_{AB}\,\ket{q}_{A}\,
  \hat D_Q^{B\dagger}(q).
  \label{eq:map-sum}
\end{equation}

\subsection{Equality of the Two Maps}
Comparing \eqref{eq:map-hom} and \eqref{eq:map-sum}, we conclude
\begin{equation}
  \boxed{\;
    \mathcal E_{\mathrm{hom+ff}}[\rho_{AB}]
    \;=\;
    \mathcal E_{\mathrm{SUM+tr}}[\rho_{AB}]
  \;}
  \qquad \forall\,\rho_{AB}.
\end{equation}
Hence, ``homodyne\,$+$\,feed-forward (with the record discarded)'' and ``SUM-then-trace'' implement operationally identical CPTP maps on subsystem $B$.

\section{Details on baseline protocols and their transduction rate.}

\begin{figure}[H]
    \centering
    \includegraphics[width=0.99\linewidth]{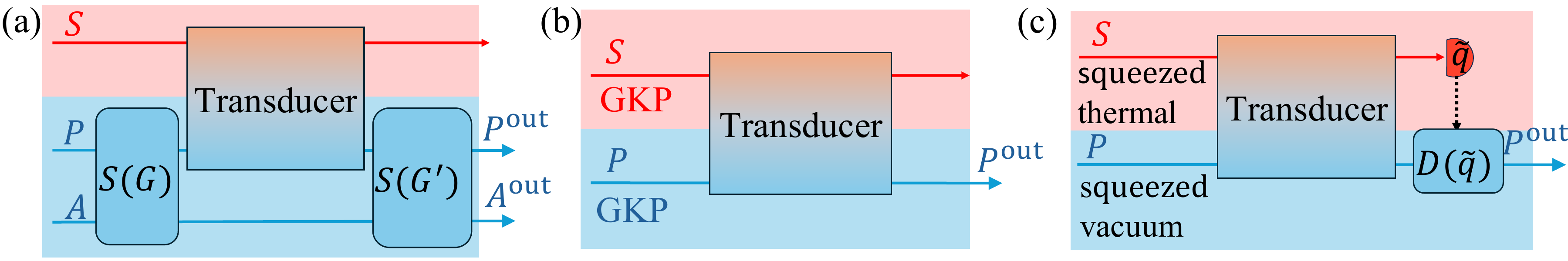}
    \caption{
Baseline transduction strategies. 
(a) Intraband entanglement–assisted protocol~\cite{shi2024overcoming}: a microwave probe $P$ and ancilla $A$ are entangled via a two-mode squeezer $S(G)$, interact with the optical signal $S$, and are processed by an antisqueezer $S(G')$.
(b) GKP-assisted protocol~\cite{Wang2025Passive}: both $S$ and $P$ are prepared in GKP states prior to the transducer. 
(c) Adaptive quantum transduction protocol~\cite{Zhang2018Quantum}: the signal mode $S$ is prepared in a squeezed-thermal state and the probe mode $P$ in a squeezed-vacuum state prior to the transducer, followed by homodyne detection in $S$ and feedforward displacement in $P$ conditioned on the measurement outcome $\tilde{q}$.
}
    \label{fig:baseline_scheme}
\end{figure}

\subsection{Nonadaptive QT}

\subsubsection{Direct QT}

For a direct QT scenario, the quantum channel for QT is a pure loss channel $\calL_\eta$ with efficiency $\eta$. Therefore, we directly adopt the quantum capacity $Q(\calL_\eta)$ specified in Eq.~\eqref{eq:Q_pure_loss_energy_constraint} to provide a benchmark.

\subsubsection{TMS-EA QT}
Entanglement assistance can enhance bosonic communication by correlating the probe and an ancillary mode prior to transduction. The intraband entanglement-assisted protocol~\cite{shi2024overcoming}, illustrated in Fig.~\ref{fig:baseline_scheme}(a), employs two-mode squeezing before the transducer and a corresponding decoding operation afterward, yielding an effective pure-loss channel with enhanced transmissivity 
\be 
\eta_{\mathrm{EA}}=1/[{1+(1-\eta)/(\eta G)}],
\ee  
where $G$ is the squeezing gain. This protocol represents the state-of-the-art among schemes that rely solely on entanglement assistance without feedforward or non-Gaussian encoding.

Similar to the direct QT case, we adopt the quantum capacity $Q(\calL_{\eta_{\mathrm{EA}}},n)$ as in Eq.~\eqref{eq:Q_pure_loss_energy_constraint}  to provide a benchmark.

\subsubsection{GKP-QT}
For a bosonic pure-loss channel, the quantum capacity is zero whenever $\eta < 1/2$. Remarkably, this threshold can be bypassed by modifying the environment state. Wang \textit{et al.}~\cite{Wang2025Passive} demonstrated that preparing both the system and environment in suitable GKP states allows perfect error correction of loss, yielding nonzero capacity for arbitrarily small~$\eta$ in the infinite–energy limit.
We therefore use this constrained GKP strategy as a non-adaptive and no-entanglement-assistance benchmark.

For the optical registers $(R,S)$, we parameterize the input as an entangled approximate GKP state,
\begin{equation}
\ket{\psi}_{RS} \sim \sum_{i=1}^{d_1} a_i e^{i\theta_i}\, \ket{i}_R \ket{(d_1,i)}_{\mathrm{GKP},S},
\end{equation}
where $\ket{i}$ denotes the $i$th Fock-basis state and $\ket{(d_1,i)}_{\mathrm{GKP},S}$ is a finite-energy approximation of a GKP qudit with logical index~$i$, parameterized as in Eq.~(\ref{eq:GKP_finite_n_general}). We take $\ket{(d_2,i)}_{\mathrm{GKP},P}$ as the input state in mode $P$. For each choice of $(d_1,d_2)$, we optimize the parameters of the input states to maximize the coherent information, subject to the energy constraints $\langle \hat n_S\rangle=\langle \hat n_P\rangle=2$. We report the best performance over $d_1\in\{2,\ldots,6\}$ and $d_2\in\{1,\ldots,5\}$.

\subsection{Adaptive-QT}
Adaptive feedforward offers a distinct mechanism for enhancing transduction by conditioning operations on measurement outcomes. The adaptive quantum transduction (AQT) protocol introduced in Ref.~\cite{Zhang2018Quantum}, shown in Fig.~\ref{fig:baseline_scheme}(c), injects squeezed ancilla states, performs homodyne detection on the signal output, and applies a conditional displacement to the probe mode. This Gaussian adaptive strategy has been shown to substantially improve channel capacity under realistic noise and serves as the primary benchmark for adaptive protocols.

As shown in Ref.~\cite{Zhang2018Quantum}, when the ancilla (mode $P$) is prepared in a finite-squeezing squeezed-vacuum state, the effective mapping from the input in mode $S$ to the output in mode $P$ reduces to a classical additive-noise Gaussian channel, denoted by $\mathcal{A}_{\sigma^2}$. A standard single-letter achievable rate for this channel is given by the coherent information evaluated on a thermal input state.  Accordingly, in our coherent-information benchmark for the AQT protocol, we initialize the environment ancilla in a squeezed-vacuum state and take mode $S$ to be a squeezed thermal state. Since different combinations of squeezing and thermal occupation can satisfy the same energy constraint, we optimize over both parameters to maximize the coherent information.

\section{GKP feature of optimal states in non-adaptive VQT}

To verify that the optimal states of non-adaptive VQT scheme in the small-$\eta$ regime are indeed GKP-like, we compute their fidelity with finite-energy GKP code states. 
Since the optimized states in modes $S$ and $P$ need not be pure after the encoding optimization, we compare them against a family of parametrized GKP-diagonal mixed states,
\begin{equation}
\rho_{\mathrm{GKP}}(d)\equiv \sum_{i=1}^{d} p_i \, \ket{(d,i)}_{\mathrm{GKP}}\bra{(d,i)}
\end{equation}
and maximize the fidelity over $\{p_i\}$ and $d\in\{1,\ldots,10\}$. 
In the optimization, we ensure that every $\ket{(d,i)}_{\mathrm{GKP}}$ has the same lattice structure (same $\hat R(\phi_2)\,\hat S(r)\,\hat R(\phi_1)$)  as in Eq.~(\ref{eq:GKP_finite_n_general})

\begin{figure}[H]
\centering
\includegraphics[width=0.95\linewidth]{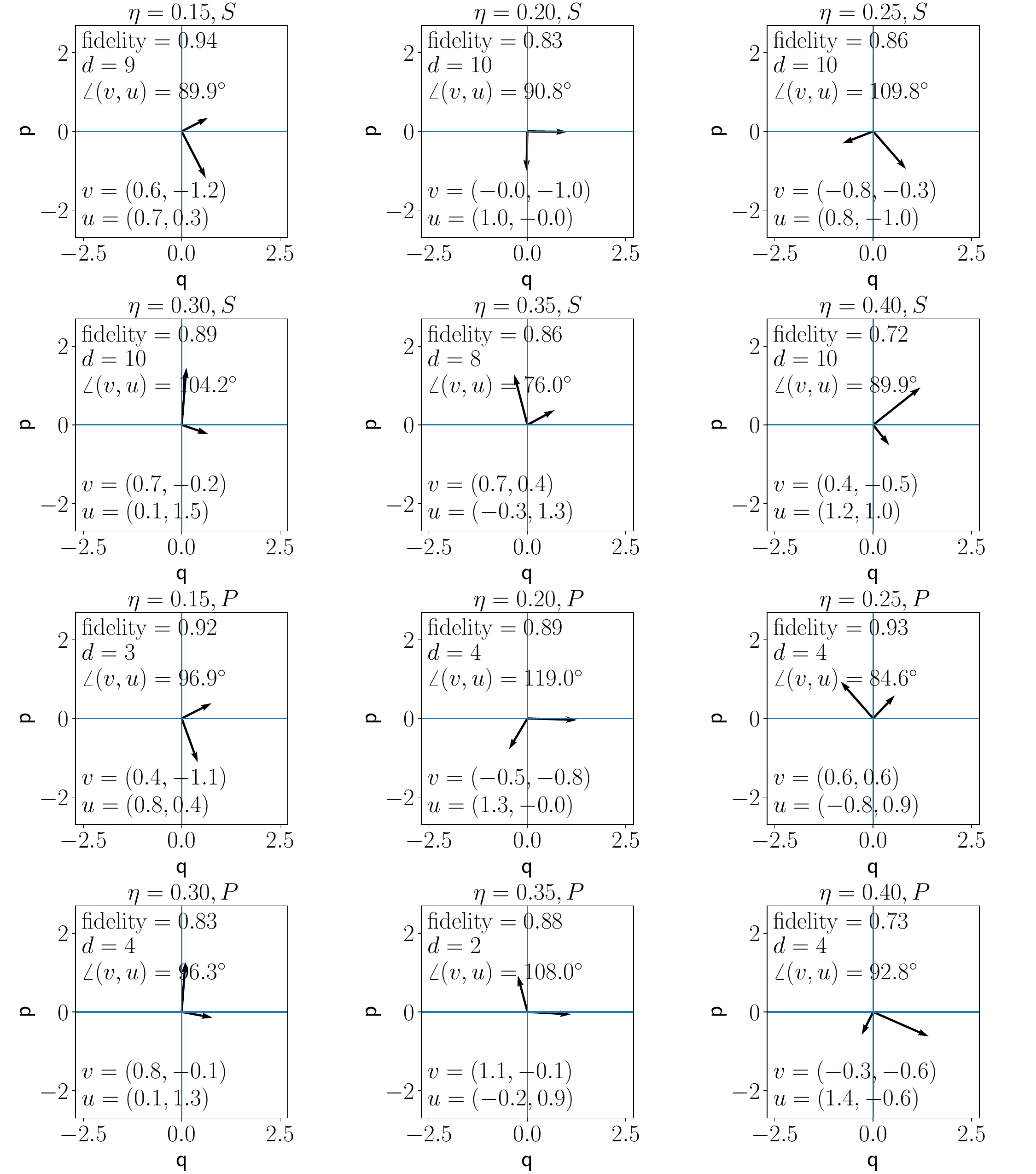}
\caption{
Optimal hexagonal-lattice generator vectors in normalized orientation matrix (Eq.\ref{eq:orientation_matrix}) for different  $\eta\in\{0.15,0.20,0.25,0.30,0.35,0.40\}$.
The first two rows are mode $S$ states and last two rows are mode $P$ states. 
In each panel, the two basis vectors $\bm{v}$ and $\bm{u}$ (drawn from the origin in the $(q,p)$ plane) are computed from the best-fit lattice parameters that maximize the fidelity over the scanned grid.
The maximal fidelity,  the corresponding optimal logical dimension $d$, angle $\angle(\bm{v},\bm{u})$ and the coordinates of $\bm{v}$ and $\bm{u}$ are reported in the annotations.}
\label{fig:GKP_state_match}
\end{figure}

The resulting maximal fidelities, reported in Fig.~\ref{fig:GKP_state_match}, lie in the range $0.72$--$0.94$ across the small-$\eta$ region. 
This constitutes a strong GKP signature, particularly under an energy constraint corresponding to $n=2$, where even ideal finite-energy square lattice GKP states exhibit moderate stabilizer expectation values, e.g.,
\begin{align}
{}_{\mathrm{GKP}}\bra{(2,0)} e^{i 2\sqrt{\pi} \hat{q}} \ket{2,0)}_\mathrm{GKP} &\approx 0.55,\\
{}_{\mathrm{GKP}}\bra{(2,1)} e^{i 2\sqrt{\pi} \hat{q}} \ket{2,1)}_\mathrm{GKP} &\approx 0.54.
\end{align}
We therefore conclude that, in the small-$\eta$ regime, the optimized states converge toward finite-energy GKP code states.
In addition, the majority of the matching GKP states have rectangular lattice, with a few having hexagonal lattice. 

\bibliographystyle{apsrev.bst} 
\bibliography{ref}

\end{document}